\documentstyle[12pt]{article}

\begin{document}
\input{epsf.sty}
\baselineskip 15pt

\title{VARIATIONS ON THE THEME OF THE GREENBERGER-HORNE-ZEILINGER
  PROOF}
\author{ Lev Vaidman}
\date{}
\maketitle

\begin{center}
{\small \em School of Physics and Astronomy \\
Raymond and Beverly Sackler Faculty of Exact Sciences \\
Tel Aviv University, Tel-Aviv 69978, Israel. \\}
\end{center}

\vspace{1.2cm}
\begin{abstract}
{\small  Three arguments based on the Greenberger-Horne-Zeilinger (GHZ) proof
  of the nonexistence of local hidden variables are presented. The first
  is a description of a simple game which a team that uses the GHZ
  method will always win. The second uses counterfactuals in an attempt
  to show that quantum theory is nonlocal in a stronger sense than is
  implied by the nonexistence of local hidden variables and the third
  describes peculiar features of time-symmetrized counterfactuals in
  quantum theory.}
\end{abstract}

\vskip 1.cm \noindent
{\bf  1. Introduction. ~~}
\vskip .2cm

Daniel Greenberger has uncovered numerous miracles of the quantum world.
Reading his work on quantum experiments with neutrons \cite{G-ne} lead
me to adopt a revolutionary view on the reality of our universe
\cite{V-mwi}. But another of his discoveries, the Greenberger, Horne,
Zeilinger (GHZ) nonlocality proof \cite{GHZ} influenced not just my view on
quantum reality, but the views  of many thousands of people. I myself
have  used this work
to explain the power of quantum mechanics to hundreds of students as
well as to many friends.
 In this paper I will discuss three arguments
based on the GHZ work.

 The first is a  version of the GHZ argument which can
 convert   laymen into admirers
of quantum theory by showing the miraculous power of quantum
theory. This is a combination of Mermin's presentation \cite{Mermin}
of the GHZ ``paradox'' and a story I heard from my students who took
a course ``Paradoxes in Quantum Probability'' by Boris Tsirelson in
Tel-Aviv University \cite{Boris}. Mermin translated the GHZ result to peculiar
correlations between the outcomes of some  simple operations. Tsirelson
constructed a certain gambling game for which a quantum team has an
advantage relative to any ``classical'' team by using the
setup of the original Bell inequalities paper \cite{Bell}. I have combined
these two works, suggesting a gambling game based on Mermin's
realization of the GHZ idea.  This argument is presented in Section
2. Section 3 is devoted to a discussion in which this  game is
considered as a method for obtaining an
experimental proof of nonlocality of quantum theory. The discussion
includes also some speculations about possible local hidden variable
theories which can explain experiments with non-ideal detectors.

The second argument was inspired by recent polemics triggered by Stapp's
proof of the ``nonlocality'' of quantum theory \cite{Stapp}. He claimed
to show, using Hardy's setup \cite{Hardy}, the nonlocality
of quantum theory beyond the result of Bell, which is the nonexistence
of a local hidden variable theory consistent with predictions of quantum
theory. I have already reflected on this subject once \cite{V-cfpsa}
using an analysis of the EPR state of two spin$-{1\over 2}$ particles,
but now I think that the GHZ setup is a better testing ground for
Stapp's claim.  In Section 4 I present an argument inspired by Stapp's
work, based on the GHZ configuration, which shows a contradiction
between certain counterfactual statements.  However, in the next section
I present my scepticism about the whole program. At this moment I cannot
define for myself what is meant by a ``local'' theory which is not a
local hidden variable theory.  Therefore, for me, all that is shown in
Stapp's and my proofs is that something which I am unable to define does
not exist. Thus, the significance of the ``proofs'' is not clear to me.

The last argument, presented in Section 6, is also related to
counterfactuals. Here I want to bring attention to an analysis based
on the argument by Clifton, Pagonis and Pitowsky \cite{Clif} about the
question of existence of a realistic relativistically-covariant quantum theory. I
do not think that one can prove the nonexistence of a covariant
realistic quantum theory \cite{V-er}, but the example allows one to
show a surprising property of time-symmetrized ``elements of reality''
which I suggest be defined in terms of certain counterfactuals.  I show that
such elements of reality do not obey the product rule. Clifton et al.
preferred not to give up the product rule and in this way they arrived
at a contradiction with the existence of a realistic covariant  quantum theory.

I conclude by explaining how the analysis of the GHZ work lead me to
accept the bizarre picture of quantum reality given by the Many Worlds
Interpretation \cite{Everett} according to which all
that we see around us is only one out of numerous worlds which all
together comprise the physical universe described by quantum theory.

\vskip 1.cm \noindent
{\bf  2. How to win a game using quantum theory. ~~}
\vskip .2cm

When I have a conversation with a friend who is not a physicist and I
want to show her the miraculous power of quantum theory I start with a
seemingly innocent puzzle.

 I present the following game for a team of three players.  The players
are allowed to make any preparations before they are taken to three
remote locations $A, B$ and $C$. Then, at a certain time $t$, each
player is asked one of two possible questions: ``What is $X$?''  or
``What is $Y$?'' (see Fig. 1).  Each  player must give an answer which is
also limited to only two possibilities: ``$1$'' or ``$-1$''. They have
to give their answer quickly, i.e., before the time they might receive a
message sent by another player after the time $t$.

According to the rules of the game, either all players are asked the
$X$ question, or only one player is asked the $X$ question and the
other two are asked the $Y$ question. The team wins if the product of
their three answers is $-1$ in the case of three $X$ questions and
is $1$ in the case of one $X$ and two $Y$ questions. My friend is asked what
should the team do in order to win for sure.

\vskip 1.7cm
\epsfxsize=9.5cm
 \centerline{\epsfbox{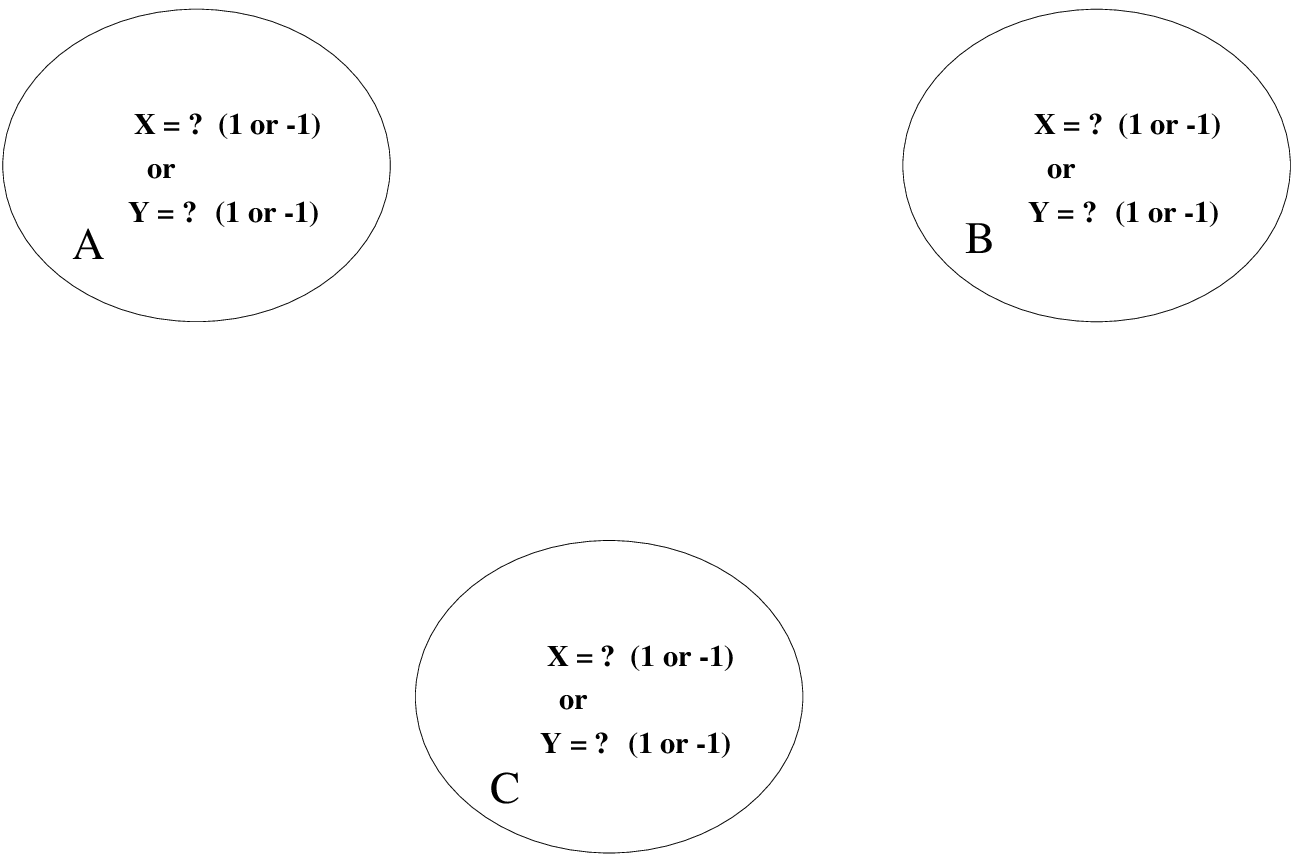}}
\vskip .2cm
\noindent
{\bf Figure 1.~ A game which only quantum team can always win.}~ Three separated
players should simultaneously provide a value for $X$ or $Y$ which might
be 1 or -1 such that the product of the values the players give will
satisfy Eq. (1).
\vskip .3cm

Sometimes the problem interests  friends immediately and sometimes
they are urged to work on this seemingly children's puzzle by my
promise that there is something unusual and profound in the solution. Usually, after less than
a half an hour I get an answer: ``This is impossible!''

 The most effective ``proof'' for this is as follows.  Since each player
is not able to get any message from the other players about which
questions they were asked before the time she has to give an answer, it
seems that she cannot gain anything by delaying the decision of which
answer to give for each question until the question is actually
asked. Thus, an optimal strategy should correspond to prior definite
decisions of each player which answers to give for possible
questions. But it is easy to prove that any such strategy cannot ensure
winning for all allowed combinations of questions. Indeed, if it does,
then there must be a set of answers $\{X_A, Y_A, X_B, Y_B, X_C, Y_C\}$,
where $X_A$ is the answer of the player in $A$ on question $X$, etc.,
such that the following equations are fulfilled. 
  \begin{eqnarray}
\label{XYZ}
\nonumber X_A X_B X_C & = & -1 ,\\
\nonumber X_A Y_B Y_C & = & 1 ,\\
Y_A X_B Y_C & = & 1 ,\\
\nonumber Y_A Y_B X_C & = & 1. 
\end{eqnarray}
 This however is impossible, because the product of all left
hand sides of equations (\ref{XYZ}) is the product of squares of numbers
which are $\pm 1$ and therefore it equals 1, while the product of all right
hand sides of these equation yields $-1$.

This is the shortest proof I know.  Many people just show this by going
through all possible strategies of deciding in advance the answers to
the questions.  After this elaborate exercise my friends  have a
firm belief that the task is impossible. At this stage I tell them that
using a quantum experiment this can be done. Usually at first my claim is
accepted with disbelief, but when I succeed to persuade them that this
is true, their surprise is enormous. 

The solution provided by  quantum theory \cite{GHZ,Mermin}
is as follows.
 Each member of the team takes with her a
spin-${1\over 2}$ particle. The particles are prepared in a correlated
state (which is usually called the GHZ state): 
\begin{equation}
\label{GHZ}
|GHZ\rangle = {1\over \sqrt
  2}{\Large (}|{\uparrow}_z\rangle_A|{\uparrow}_z\rangle_B|{\uparrow}_z\rangle_C - 
|{\downarrow}_z\rangle_A|{\downarrow}_z\rangle_B|{\downarrow}_z\rangle_C{\Large
)} . 
\end{equation}
Now, if a member of the team is asked the $X$ question, she measures
$\sigma_x$ and gives the answer which she obtains in this experiment. If
she is asked the $Y$ question, she measures $\sigma_y$ instead. Quantum
theory ensures that the team following this strategy {\em always}
wins. Indeed, a straightforward calculation shows that the measurements
on the system of three spin-${1\over 2}$ particles prepared in the state
(\ref{GHZ}) fulfill the following relations:
\begin{eqnarray}
\label{sigmaXYZ}
\nonumber \{{\sigma_A}_x\} \{{\sigma_B}_x\} \{{\sigma_C}_x\} & = &  -1 ,\\
\nonumber \{{\sigma_A}_x\} \{{\sigma_B}_y\} \{{\sigma_C}_y\} & = & 1 ,\\
\{{\sigma_A}_y\} \{{\sigma_B}_x\} \{{\sigma_C}_y\} & = & 1 , \\
\nonumber \{{\sigma_A}_y\} \{{\sigma_B}_y\} \{{\sigma_C}_x\} & = & 1 .
\end{eqnarray}
\noindent
Here $\{{\sigma_A}_x\}$ signifies the outcome of the measurement  of
$\sigma_x$ by  the player  in $A$, etc. Let us show, for example, that the
first equation  is true. In the spin $x$ bases for all particles the
GHZ state (which in (\ref{GHZ})  is given in the spin $z$ bases) is
\begin{eqnarray}
\nonumber
|GHZ\rangle = {1\over 
  4}[(|{\uparrow}_x\rangle_A +
|{\downarrow}_x\rangle_A)(|{\uparrow}_x\rangle_B +
|{\downarrow}_x\rangle_B)(|{\uparrow}_x\rangle_C +
|{\downarrow}_x\rangle_C)\\
\nonumber - (|{\uparrow}_x\rangle_A -
|{\downarrow}_x\rangle_A)(|{\uparrow}_x\rangle_B -
|{\downarrow}_x\rangle_B)(|{\uparrow}_x\rangle_C -
|{\downarrow}_x\rangle_C)]\\
= {1\over 
  2}(|{\uparrow}_x\rangle_A
|{\uparrow}_x\rangle_B|{\downarrow}_x\rangle_C
 + |{\uparrow}_x\rangle_A
|{\downarrow}_x\rangle_B|{\uparrow}_x\rangle_C\\
\nonumber +
|{\downarrow}_x\rangle_A
|{\uparrow}_x\rangle_B|{\uparrow}_x\rangle_C +
|{\downarrow}_x\rangle_A
|{\downarrow}_x\rangle_B|{\downarrow}_x\rangle_C).
\end{eqnarray}
Therefore, we see explicitly that the GHZ state is a superposition of states
for each of which $\{{\sigma_A}_x\} \{{\sigma_B}_x\} \{{\sigma_C}_x\} =
-1$.

In the quantum solution of the problem the players do not decide in
advance the answers they'll give for each out of the two possible
questions. In the ``proof'' of the impossibility of this task
presented above it was erroneously assumed that  delaying the
decision which answer to give until the time the question is asked
cannot help.  The assumption sounds plausible since 
relativistic causality prevented sending signals after the time the
questions were asked, but, nevertheless, the assumption is wrong
because it does not take into account unusual correlations which
quantum objects can exhibit.

\vskip 1.cm \noindent
{\bf  3. The GHZ proof of nonexistence of local hidden variable
  theories.
~~}
\vskip .2cm

Performing the game with spin$-{1\over 2}$ particles as described above
might serve as an experimental proof of the nonexistence of local hidden
variables which produce agreement with experiment. Suppose, however,
that Nature {\em is} governed by local hidden variables. This means that
the results $\{{\sigma_A}_x\}, \{{\sigma_A}_y\}, \{{\sigma_B}_x\}$, $
\{{\sigma_B}_y\}, \{{\sigma_C}_x\}, \{{\sigma_C}_y\}$ exist prior to
their measurements and, therefore, there are answers $\{X_A, Y_A, X_B,
Y_B, X_C, Y_C\}$ prior to the time the questions are asked. These
answers can fulfill at most three out of four equations (\ref{XYZ}) and,
therefore, in the game in which the four question patterns are chosen
with equal probability, on average there will be at most a 75\% success
rate of any team playing this game.  If a team shows on average a higher
result, this is an experimental proof that Nature is not governed by
local hidden variables.

The most common objection to  experiments showing the nonexistence
of local hidden variables is that, due to limited efficiency of
detectors, a significant fraction of particles is lost. If every member of
our team  using the quantum strategy described above makes a random choice when she does not detect  the spin component of
the  particle, then the success rate of the team is 
 \begin{equation}
{\rm Prob}(win) =  {\rm Prob}(all~detected) + {1\over 2} (1-  {\rm
  Prob}(all~detected)) .
\end{equation}
Therefore, to ensure that Prob({\it win}) $>$ 0.75 (the maximal value
achievable by a classical team), so that the nonexistence of the hidden variables is shown, it is enough that Prob({\it
  all~detected}) $>$ 0.5, i.e., that the efficiency of each detector is
bigger than $0.5^{1\over 3} \sim 0.8$. Note  much more sophisticated
 studies of limitations posed by detector efficiencies  by
Greenberger, Horne, Zeilinger,  and Shimony \cite{GHZS} and  Larsson \cite{Lars}.

It is interesting to think about possible local hidden variable theories
that can produce results equivalent to those of quantum theory.  For the
experiment with limited efficiency  detectors in which Prob({\it all~detected})
$<$ 0.5 such a theory exists \cite{Phil}. The spin$-{1\over 2}$
particles of a GHZ triple, when they were locally created, and then
moved to their locations $A$, $B$, and $C$, carry with them
``instruction kits'' how to respond to various measurements.  There are
instructions for every possible direction of the spin measurement which
are: ``up'', ``down'', or ``be not detected''.  Indeed, it is always
possible to construct instructions how each spin is to respond to
$\sigma_x$ and $\sigma_y$ measurements such that two of equations
(\ref{sigmaXYZ}) are fulfilled and  just one spin$-{1\over 2}$ particle
has an instruction ``be not detected'' for one of the directions and
therefore the other two equations (\ref{sigmaXYZ}) are not tested.  In
this situation, on average, in half of the measurements there will be
triple detections for all of which the results are in accordance with
(\ref{sigmaXYZ}).  (Note that in this model in every run at least two
particles are detected, i.e., no single or null detections can
occur. This is not expected according to quantum theory which does not
have predictions about correlations between detection or not detection
of particles.)

The naive local hidden variable theory proposed above fails to explain
the GHZ correlations in a more sophisticated setup. Roughly speaking I
propose ``teleportation'' \cite{tele} of the states of the GHZ particles
to three  particles in even more remote locations and testing the GHZ correlations on
the remote particles. Quantum theory predicts such correlations but the
naive local variables theory can not. Indeed, the
remote particles have no ``common cause'', they were not locally
created, and therefore they cannot carry with them instruction kits
created in a single location. Thus, the naive hidden variable theory
cannot explain why the three remote particles exhibit quantum
correlations.

  More precisely, I propose the following experiment.  Three EPR pairs
are prepared as for the teleportation of the states of the GHZ particles
as shown in Fig. 2. Then, simultaneously, three Bell operator
measurements are performed on the pairs of spin$-{1\over 2}$ particles
consisting from one GHZ particle and one particle from the EPR pair at
$A$, $B$ and $C$. At
the {\em same} time spin-$x$
or spin-$y$ components of the three  particles  at
$A'$, $B'$, and $C'$,  are measured according to the usual GHZ rule:
either three $x$ or just one $x$ and two $y$ components are  measured.

In the original teleportation scheme \cite{tele}, the outcome of the
Bell measurement is transmitted to the remote particle and a $\pi$
rotation around one of the three axes (or no rotation) is performed
according to that outcome. In the proposed experiment, no message is
sent and no rotation is performed, because the spin component
measurement is performed before the message about the result of the Bell
measurement could have arrived.  The significance of the Bell
measurement is that it makes the connection between the result of the
spin measurement on the remote particle of the EPR pair and the result
of possible direct measurement of the same spin component of the GHZ
particle. For two out of four possible results of the Bell measurement
(corresponding to no rotation and the rotation around the axis of the
direction of the spin component measurement) measurement of the remote
particle is identical to the direct measurement performed on the GHZ
particle. For two other results, the outcome of the mea-
\break
\vskip .3cm
\epsfxsize=9.5cm
 \centerline{\epsfbox{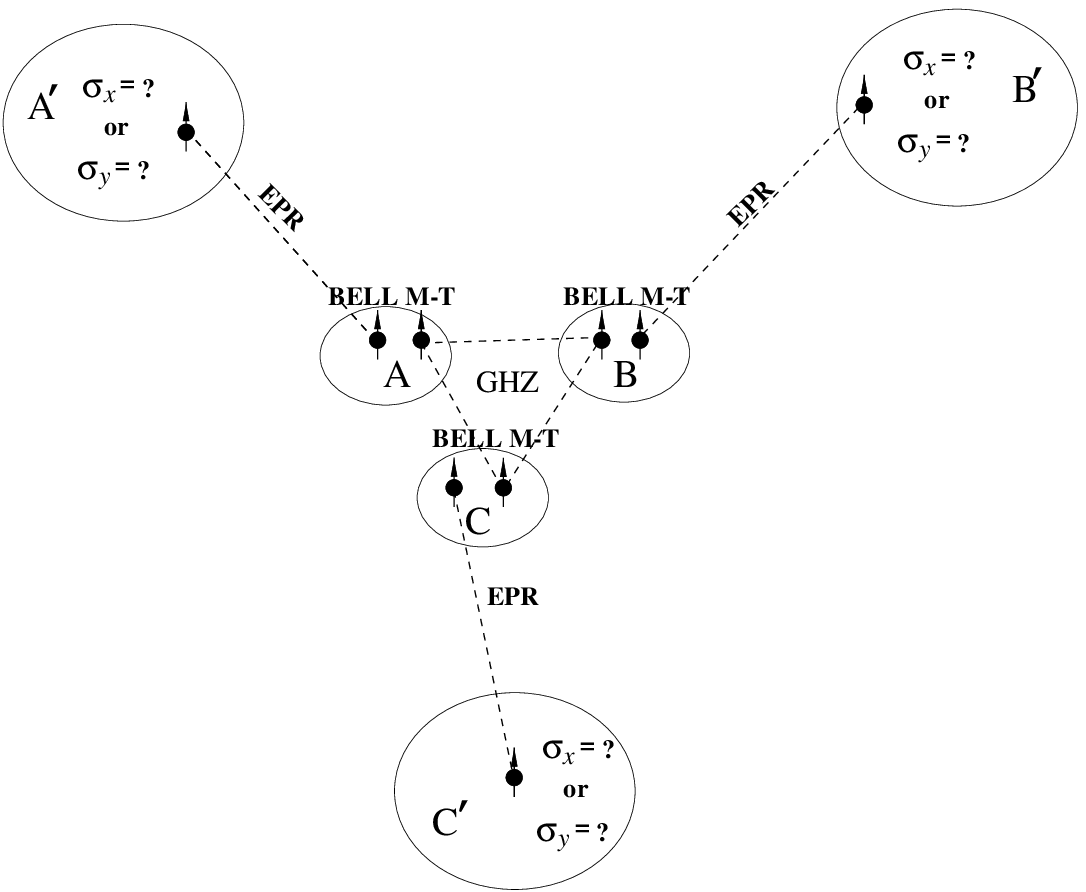}}
\vskip .1cm
\noindent
{\bf Figure 2.~ A proposal for experiment for ruling out ``naive'' local
hidden variable theories.}~ There is a free choice of measurements in
$A'$, $B'$ and $C'$ and  fixed (Bell) measurements in $A$, $B$ and
$C$. Quantum theory predicts certain correlations between the outcomes
of these measurements, but since the particles in $A'$, $B'$ and $C'$
had no common origin they cannot carry instruction kits to produce such
correlations.

%\vskip .9cm

\noindent
surement on the
remote particle is ``flipped'' relative to the measurement performed 
on
the GHZ particle.  After taking into account these flips the
correlations between the outcomes of the spin measurements on the remote
particles turn out to be the GHZ correlations.

Since simultaneous detection of 9 particles is necessary for the case
when the correlation can be tested there will be only a small fraction
of such events.  But in this setup the remote particles cannot carry
with them instruction kits created at a single place because the
particles never were together at a single location. Therefore, it is
not clear how a hidden variable theory can produce {\em any}
correlation between the outcomes of the measurement of the remote
particles.  I do not have a proof that a more complicated local hidden
variable theory according to which the instructions for each particles
will involve not just dependence on the experimental setup of the
detectors, but also a dependence on the local hidden variables of the
other particles, cannot be created.  However, it seems to be highly
improbable to have such a theory which would not look artificial.

The GHZ proof is the most clear and persuasive proof of nonexistence
of local hidden variables. In the GHZ example we have {\em
  perfect} correlations which cannot be reproduced by any local hidden
variable theory.  This, however, does not mean that the GHZ-type
experiment is the best candidate for experimental proof of the
nonexistence of local hidden variables. There are serious
technological difficulties with experiments involving GHZ type
correlation. (Note, however, that the entanglement swapping
(teleportation), discussed in the modified test of local hidden variables
above, has been performed in the laboratory \cite{swap}.)
Today, the best choices for experimental proof of the nonexistence of
local hidden variables are still experiments based on the two-particle
quantum correlations, either via the original Bell proposal or through
ideas of Hardy \cite{Hardy} using which one can construct another game
for a two-player team such that the success in the case of the
existence of a local hidden variable theory must be less than the
success predicted by quantum theory.

\vskip 1.cm \noindent
{\bf  4. The Stapp nonlocality argument applied to the GHZ setup.~~}
\vskip .2cm

Recently Stapp \cite{Stapp}  suggested that 
the nonlocality of quantum theory goes beyond the nonexistence of
a local hidden variable theory. In order to prove this Stapp applied {\em counterfactual}
reasoning to  Hardy's setup  \cite{Hardy}.
I shall use here counterfactual arguments about the GHZ setup inspired
by Stapp's  work.

The general locality principle is 
\noindent\begin {quotation}
\noindent $\cal L$:~ {\em Action at a space-like
separated region does not change the outcome of a local
measurement.}
\end{quotation}
My understanding of $\cal L$ 
 in the context of recent
publications on this subject  is as follows. Assume that in the actual world a 
quantum experiment has a certain outcome. Then, in a counterfactual
world which differs from the actual one prior to the measurement only
in some actions performed in a space-like separated region, the outcome of the
measurement should be the same. I believe that this is, essentially,  the assumption
``LOC1'' of
Stapp's paper. I suggest  extending the meaning of the locality
postulate from comparing actual and counterfactual worlds to
comparison between two counterfactual worlds. Consider two
counterfactual worlds in which a certain measurement is
performed. Assume that the two counterfactual worlds are identical
prior to the measurement except for some actions performed in 
space-like separated regions. Then the outcomes of the measurements in
the two counterfactual worlds must be the same.

Consider the GHZ setup of three spin$-{1\over 2}$ particles located
in three space-like separated regions. Assume that in the actual world the
outcomes are:
  \begin{equation}
\label{out}
\{{\sigma_A}_x\} =1,~~~  \{{\sigma_B}_x\} =1, ~~~~\{{\sigma_C}_x\} = -1.
\end{equation}
Now consider three counterfactual worlds 
%\indent
 \begin{quotation}
\indent 
CFW1: ${\sigma_A}_x,~ {\sigma_B}_y$ and ${\sigma_C}_y$ are
measured.\\
\indent CFW2: ${\sigma_A}_y,~ {\sigma_B}_x$ and ${\sigma_C}_y$ are
measured.\\
\indent CFW3: ${\sigma_A}_y,~ {\sigma_B}_y$ and ${\sigma_C}_x$ are
measured.
\end{quotation}
\noindent Since in the actual and counterfactual worlds which  differ only by
actions in regions which are space-like separated from a certain
space-time location  the outcomes of local measurements in this location should be the
same, we may conclude that the results of ${\sigma}_x$ measurements in the
counterfactual worlds must be identical to those in the actual
world, given in (\ref{out}).
 Therefore, in order to fulfill (\ref{sigmaXYZ}) we must have:
  \begin{eqnarray}
\label{incon1}
\nonumber \{{\sigma_B}_y^{CFW1}\} = \{{\sigma_C}_y^{CFW1}\} ,\\
\{{\sigma_A}_y^{CFW2}\} = \{{\sigma_C}_y^{CFW2}\} ,\\
\nonumber \{{\sigma_A}_y^{CFW3}\} \neq \{{\sigma_B}_y^{CFW3}\}.
\end{eqnarray}
 From the extended argument which considers two counterfactual
worlds we can deduce that the results of ${\sigma}_y$ measurements
must  yield identical outcomes  for each pair of the
counterfactual worlds. Therefore, we obtain:
 \begin{eqnarray}
\label{incon2}
\nonumber \{{\sigma_C}_y^{CFW1}\} = \{{\sigma_C}_y^{CFW2}\} ,\\
\{{\sigma_B}_y^{CFW1}\} =  \{{\sigma_B}_y^{CFW3}\} ,\\
\nonumber \{{\sigma_A}_y^{CFW2}\} = \{{\sigma_A}_y^{CFW3}\}.
\end{eqnarray}
It is easy to see, however, that equations (\ref{incon1}) and (\ref{incon2}) are
inconsistent. This inconsistency shows that predictions of quantum
theory lead to failure of the general locality
principle $\cal L$, i.e., that the quantum theory is nonlocal.

My argument might be attacked on the grounds that my extension of the
locality principle to comparison of two counterfactual worlds instead of
the actual world and a counterfactual world is not justified.  In the
framework of a hidden variables theory the justification of my step is
trivial, but without hidden variables the justification is not clear. I
contend that without hidden variables even the unextended locality
principle (such as Stapp's LOC1) is not clear. Therefore, it is not
clear that my proof or even the original Stapp's proof shows anything
new about the nonlocality of quantum theory. I shall discuss this issue
in the next section.

\break

\vskip 1.cm \noindent
{\bf  5. Do the Stapp    arguments show the nonlocality of
  quantum theory beyond the nonexistence of local hidden
variables?~~}
\vskip .2cm

It seems to me that the answer to the question in the title of this
section is negative, but I certainly cannot prove it because the truth
of this statement depends crucially on the meaning of ``nonlocality''
which differs widely among physicists and philosophers.  What I want
to present here are some arguments relevant to {\em my} understanding
of nonlocality in this context formed after reading an illuminating
discussion by Mermin \cite{Mermin1} of the previous version of Stapp's
nonlocality argument.  The title of Mermin's paper is: ``Can You Help
Your Team Tonight by Watching on TV? More Experimental Metaphysics
from Einstein, Podolsky, and Rosen.'' The word ``help'' might be too
strong: the question of nonlocality, as I understand it, is: ``Can you
{\em change} an outcome of the game by an action in a space-like
separated region?'' A one-sentence summary of my arguments below is:
for counterfactual propositions in the EPR setup the word ``change''
is meaningless unless the existence of hidden variables is assumed.

Stapp's paper \cite{Stapp} generated a very intensive polemic
including critical analysis by Unruh, Mermin, Finkelstein, Griffiths,
Shimony and myself \cite{Unruh,Mermin2,Fink,Griff,Shim,V-cfpsa} and
answers by Stapp \cite{Stapp1,Stapp2}. The ultimate goal of the
project, as I understand it, is to show that quantum theory invariably
leads to the failure of locality principle $\cal L$.  Two relevant
results are well known.
\noindent
\begin{quotation}
\phantom{p}

\noindent
(i) According to quantum
theory, action at a space-like separated region does not change the {\em
probability} of an outcome of a local measurement.

\noindent
(ii) If the outcomes of quantum measurements are governed by hidden
variables, then there is an action at a space-like separated region
which {\em does} change the outcome of a local measurement.
\end{quotation}
 I will argue that there is no meaning
for  $\cal L$  beyond  (i) and (ii).

If we assume the nonexistence of hidden variables, then the outcome
which does not have probability 1 is uncertain prior to the
measurement. Therefore, the only things we can compare are {\em
  probabilities} for an outcome so, strictly speaking, $\cal L$ is
meaningless. $\cal L$ can be made meaningful only if we read ``the outcome'' in $\cal L$ as ``the
probability of the outcome'': then the meaning of $\cal L$ is (i) and
$\cal L$ is true.

Under the opposite assumption of existence of hidden variables the
meaning of $\cal L$ is (ii) and $\cal L$ is false. But this is nothing new, the
nonexistence of local hidden variables is known, and a nonlocal hidden
variable theory means exactly the negation of $\cal L$ for some
measurements.

The definition of hidden variables is that the outcome of any
experiment is known prior to experiment and the definition of {\em local}
hidden variable is that the outcome of any local experiment is known
prior to the experiment. (Frequently, the concept of {\em
  contextuality} is introduced in the discussion of this issue. This
should not change the argument provided that the context of a local
experiment is also local, i.e., relates only to the location of the
experiment.)  

A typical situation for which $\cal L$ is applied in the Stapp type
proofs is that a local quantum measurement has several possible
outcomes and a particular one takes place in the actual world. The
locality principle says that in a counterfactual world, which differs
from the actual world in some action performed in a space-like
separated region, the outcome should not be {\em different}. Now, a
counterfactual world should be as close as possible to the actual
world.  Since the question is about the result of a local measurement,
the counterfactual world should be as close as possible to the actual
world {\em prior} to the measurement. But, since a hidden variable is
not assumed, in the actual world prior to the measurement the outcome
is not known. But if the outcome is not known, then the word ``{\em
  different}'' in the sentence above describing the locality principle
becomes meaningless: different from what? There is nothing to compare
with unless a hidden variable which determines the outcome is assumed.
Therefore, it seems that there is no meaning for nonlocality of
quantum theory beyond the nonexistence of local hidden variables.

Let me quote  Stapp's recent paper \cite{Stapp1}:
\begin{quotation}
 With fixed initial 
conditions one can, by making a change in the Lagrangian in a small 
space-time region, shift from the actual world to a neighboring possible
world, and prove that the effects of this change are confined to times that
lie later than the cause in every Lorentz frame. The change in the 
Lagrangian in the small region can be imagined to alter an experimenter's
choice of which experiment he will soon perform in that region.

... The question, more precisely, is this: Is it possible to maintain  
in quantum mechanics, as one can in classical mechanics, the theoretical idea 
that the one real world that we experience can be embedded in a set of 
possible worlds, each of which obeys the known laws of physics, if (1), the 
experimenters can be imagined to be able to freely choose between the 
different possible measurements that they might perform, and (2), no such 
free choice can have any effect on anything that lies earlier in time in 
some Lorentz frame. 

It was proved in  \cite{Stapp} that with a sufficiently broad definition of 
``anything'' the answer to this question is no.
\end{quotation}

In his proof Stapp considers an actual world with local measurements
in two separate locations which have certain outcomes. Then, using (1)
and (2) he shows that certain worlds with some alternative
measurements should belong to the set of possible worlds.  It seems to
me that Stapp arrived at a contradiction (which he argued implies the
nonlocality of quantum theory) by relying on a tacit assumption that
in a possible world, identical to the actual world in {\it everything}
prior to a measurement, the result of the same measurement which was
performed in the possible world must be identical to that of the
actual world.  Thus, he excludes from the set of possible worlds those
which have different results for the {\it same} measurements given
fixed initial condition. However, the quantum theory predicts {\em
  random} outcomes and therefore this exclusion is not justified.
Saying this again in the language of the previous consideration, this
exclusion fixes the outcome prior to a measurement, i.e., this
strategy assumes existence of hidden variables.

The assumption of nonexistence of hidden variables, that is, for
example, that the outcome of a spin component measurement performed on
one particle out of the EPR pair (singlet state of two separated
spin$-{1\over 2}$ particles) is not fixed prior to the measurement,
leads to some ``nonlocal influence'': measurement on one particle fixes
{\em immediately} the outcome of possible  measurement  on the
other particle. This, however, does not correspond to the failure of
$\cal L$. We cannot claim that the measurement on one particle {\it
  changes} the outcome of the measurement performed on the other
particle. The latter is {\em random} with or without measurement
performed on the first particle. The ``randomness'' is the same
irrespectively of where it was generated, in the local measurement itself,
or in the remote measurement.

Thus, it seems that without assuming hidden variables Stapp's
arguments cannot show the failure of $\cal L$.  According to my
understanding, Stapp views ``nonlocality'' as something different from
``the failure of $\cal L$'' and he obviously believes that his proof
shows something more than the existence of an instantaneous fixing of
an outcome of a remote possible measurement -- a trivial consequence
of the assumption of the nonexistence of hidden variables and the
existence of a ``free'' local agent which choses the measurement to be
performed.  Note, however, that my results are in agreement with the
claim of Shimony \cite{Shim}, who, by analyzing Stapp's proof itself,
reached the conclusion that Stapp's nonlocality proof does not go
through unless hidden variables are assumed.

\vskip 1.cm \noindent
{\bf  6. Elements of reality in the  GHZ setup.~~}
\vskip .2cm

Several years ago an interesting argument based on the GHZ setup was
advanced by Clifton et al. \cite{Clif}. They arrived at a contradiction
assuming existence of a realistic covariant description of quantum
systems. Inspired by their work I constructed time-symmetrized
``elements of reality'' in terms of which there is no contradiction.
The price of removing the contradiction is a peculiar feature of these
elements of reality: their product rule fails.

I have proposed the following {\em definition} of elements of reality
\cite{V-er}:
  \begin{quotation}
If we can infer with
certainty the result of  measuring a
physical quantity at time $t$ then, at time $t$, there exists an element of
reality corresponding to this physical quantity which has a  value equal to
the predicted measurement result.
  \end{quotation}
  This is a modification of what is usually considered to be a
  necessary {\em property} of an element of reality. The
  time-symmetrization is in the word ``infer'' which replaced the
  asymmetric term ``predict''.

Consider three particles prepared in the GHZ state at $t_1$ which were
measured later at time $t_2$ and the results (\ref{out}) were found. Now we
can consider elements of reality which are counterfactual statements about measurements at time $t$,
$t_1 < t < t_2$.
Just from the fact that the particles are prepared in the GHZ state, it
follows that
\begin{eqnarray}
\label{sigmaX-Y-Z}
\{{\sigma_A}_x {\sigma_B}_x {\sigma_C}_x\} & = & -1 , \nonumber \\
\{{\sigma_A}_x {\sigma_B}_y {\sigma_C}_y\} & = & 1 , \nonumber \\
\{{\sigma_A}_y {\sigma_B}_x {\sigma_C}_y\} & = & 1 ,  \\
\{{\sigma_A}_y {\sigma_B}_y {\sigma_C}_x\} & = & 1 . \nonumber  
\end{eqnarray} 
Note that these equations are different from (\ref{sigmaXYZ}) which
relates to the product of the outcomes of separate spin component measurements at each location, while
(\ref{sigmaX-Y-Z}) relates to the measurement of the products of
the spin components of the GHZ particles. The operators of the products commute
when applied to the GHZ state and therefore (\ref{sigmaX-Y-Z}) are not
``true'' counterfactuals: they could be called  ``conditionals'' because, in
principle, they can be measured together without disturbing each other (not
that I know how to do it). In contrast, equations  (\ref{sigmaXYZ})
are ``true''  counterfactuals
because they cannot be measured together. However, equations (\ref{sigmaXYZ})
are not exactly ``elements of reality''. (I named them ``generalized
elements of reality'' \cite{V-prep-cf} because each equation yields a
certain relation between the outcomes of several measurement and not
just the value of a single  measurement as  appears in the definition of the
element of reality.)

 Taking into account the results of the measurements
at $t_2$ given by (\ref{out}) we can derive the following list of
elements of reality related to the intermediate time $t$:
  \begin{eqnarray}
\label{sigmaX-Y}
\nonumber  \{{\sigma_A}_y {\sigma_B}_y \} = -1 ,\\
\{{\sigma_A}_y {\sigma_C}_y \}  = 1 ,\\
\nonumber \{{\sigma_B}_y {\sigma_C}_y \} = 1 .
\end{eqnarray} 
Note the difference between application of counterfactuals here and in
the previous section which allows derivation of (\ref{sigmaX-Y}). In the
present section, the counterfactual worlds have fixed both initial and
final conditions, i.e., the outcomes of the measurements at $t_1$ and $t_2$,
in contrast to the situation in previous section in which only initial
conditions were fixed.

Now we can consider the product of  equations (\ref{sigmaX-Y}):
 \begin{equation}
\label{prod}
 \{{\sigma_A}_y {\sigma_B}_y \}\{{\sigma_A}_y {\sigma_C}_y \}
 \{{\sigma_B}_y {\sigma_C}_y \} = -1.
\end{equation}
 What is peculiar here is that the outcome of the measurement  of the
 product of the operators appearing on the left hand side of of (\ref{prod})
is also known with certainty, i.e., we have an element of reality:
 \begin{equation}
\label{produ}
 \{{\sigma_A}_y {\sigma_B}_y {\sigma_A}_y {\sigma_C}_y 
 {\sigma_B}_y {\sigma_C}_y \} = \{{\sigma_A}^2_y {\sigma_B}^2_y
 {\sigma_C}^2_y \}= 1.
\end{equation}
Thus, we have shown the failure of the product rule for
time-symmetrized elements of reality. The product rule (which holds
for time-asymmetric elements of reality with ``predict'' instead of
``infer'') is: ``If $A=a$ is an element of reality and $B=b$ is an
element of reality, then  $AB=ab$ is also an element of reality''.

\vskip 1.cm \noindent
{\bf  7. Conclusions. ~~}
\vskip .2cm

I want to finish this paper by stressing again the lesson  I
learned from the GHZ proof and other works of Daniel Greenberger. As
far as I know, he himself does not draw this conclusion, but I became
a strong believer in the Many-Worlds Interpretation (MWI) \cite{Everett} of quantum theory.
Only in this framework do the difficulties of the GHZ setup  not lead
to nonlocal physical action. There is no nonlocal action on the
level of the theory of the whole physical universe which incorporates
our world and many other worlds, while we can see that inside a particular
world we obtain effectively a nonlocal action. 

According to the MWI the outcomes (\ref{out}) take place in one world
and there are, in parallel, three other worlds corresponding to
 \begin{eqnarray}
\label{sigmaX-Y-Z-}
& \{{\sigma_A}_x\} = -1, \;\;\; \{{\sigma_B}_x\} = \phantom{-}1, 
  \;\;\; \{{\sigma_C}_x\} = \phantom{-}1; \nonumber \\
& \{{\sigma_A}_x\} = \phantom{-}1, \;\;\; \{{\sigma_B}_x\} = -1, 
  \;\;\; \{{\sigma_C}_x\} = \phantom{-}1; \\
& \{{\sigma_A}_x\} = -1, \;\;\; \{{\sigma_B}_x\} = -1, 
  \;\;\; \{{\sigma_C}_x\} = -1 . \nonumber
\end{eqnarray}
In each location the only change due to the measurement
interaction is that
each spin, originally correlated with the states of the  two remote
spins, becomes correlated also
with a state of a measuring device. However, the complete local
description of each spin, its density matrix, remains unchanged. On the
other hand, ``inside'' a particular world when the particular outcomes of
 spin measurements of two particles are taken into account, the third particle
 is described by a pure state and therefore its density matrix is
 changed. Thus, no nonlocal actions take place in Nature, but
 nevertheless there is an explanation why {\em we} observe seemingly
 nonlocal phenomena.

 It is a pleasure to thank Philip Pearle, Sandu Popescu, and Abner
 Shimony for illuminating discussions and Jerry Finkelstein, Robert
 Griffiths, David Mermin, Henry Stapp, and Bill Unruh for helpful
 correspondence.  The research was supported in part by grant 614/95
 of the Basic Research Foundation (administered by the Israel Academy
 of Sciences and Humanities). Part of this work was done during the
 1998 Elsag-Bailey Foundation research meeting on quantum computation.

\end{document}